\documentclass[11pt]{article}
\usepackage{times}
\usepackage{amssymb}
\usepackage{amsmath}
\usepackage{amsthm,enumerate}
\usepackage{xcolor}
\usepackage{url,hyperref}

\usepackage{rotating} 
\usepackage{xspace}
\usepackage{array}

\newcommand{\comm}[1]{}

\newtheorem{thm}{Theorem}
\newtheorem{prop}{Proposition}
\newtheorem{lem}{Lemma}
\newtheorem{cor}{Corollary}
\newtheorem{con}{Conjecture}

\def\k{\kappa}

\title{A construction of product blocks  with a fixed block size}

\author{Sergey Bereg\thanks{
Department of Computer Science,
Erik Jonsson School of Engineering and Computer Science,
University of Texas at Dallas.}
}
\date{}

\begin{document}
\maketitle

\begin{abstract}
Let $M(n,d)$ be the maximum size of a permutation array on $n$ symbols with pairwise
Hamming distance at least $d$. 
Some permutation arrays can be constructed using blocks of certain type \cite{bmmms-arx-18} called {\em product blocks} in this paper. 
We study the problem of designing $(q,k)$-product blocks with a fixed block size $k$. 
\end{abstract}

\section{Introduction}

Recently, new bounds for permutations arrays found in \cite{bmmms-arx-18} by applying the contraction operation  
\cite{bls-cpag-18} to the groups $AGL(1,q)$ and $PGL(2,q)$ for a prime power $q$ satisfying $q\equiv 1 \pmod 3$. 
The contraction of $PGL(2,q)$ gives rise to a new problem of finding a maximum independent set in the contraction graph.
This problem reduces to another interesting problem of designing blocks satisfying some conditions.

For the sake of simplicity, consider a special case first where $q$ is a prime number.
We define {\em $q$-product blocks} as a collection $B_1,B_2,\dots,B_q$ of subsets of $\{1,2,\dots,q\}$
such that for any two elements $a\in B_i$ and $b\in B_j$ with $i<j$,
\begin{equation} \label{eq1}
(b-a)(j-i)\ne 1\pmod q.
\end{equation}
Clearly, one can identify $q$ and 0 and define the {\em $q$-product blocks} as 
a collection $B_0,B_1,\dots,B_{q-1}$ of subsets of $\{0,1,2,\dots,q-1\}$ such that
for any two elements $a\in B_i$ and $b\in B_j$ with $i<j$, Equation (\ref{eq1}) holds.
We will use this definition in the paper.

In general, when $q$ is a prime power, we define {\em $q$-product blocks}, $q$-PB for short, as a collection of $q$ blocks 
$B_s, s\in GF(q)$ (labeled by the elements of a Galois field $GF(q)$) such that (i) each block $B_s\subseteq GF(q)$, and 
(ii) for any two elements $a\in B_r$ and $b\in B_s$, 
\begin{equation} \label{eq2}
(b-a)(s-r)\ne 1\text{ in } GF(q).
\end{equation}

\begin{thm} \label{thm1}
If there exist $q$-product blocks of total size $v$ for a prime  $q=1\pmod 3$,
then $M(q,q-3)\ge (q-1)(v+q)$.
\end{thm}

In this paper we study $q$-product blocks with a fixed block size $k$, $(q,k)$-PB for short. 
In particular, we are interested in block constructions with large block size.
Let $\k(q)$ be the largest number $k$ such that there exist $q$-product blocks with block size $k$.
We also propose to study block constructions using only some number of elements of $GF(q)$, say $t$ elements of $GF(q)$. 
We call a set of $q$-product blocks a {\em $(q,k,t)$-PB} if it has block size $k$ and uses only $t$ elements of $GF(q)$.

\section{$(q,k,t)$-product blocks}

We show some bounds of $\k(q)$ and properties of $(q,k)$-PBs. 

\begin{prop} \label{l1}
{\rm (i)} 
For any prime power $q$, $\k(q)\ge 1$. For example, the blocks could be set $B_i:=\{a\}, 0\le i<q$, for some fixed $a$ in $GF(q)$.\\
{\rm (ii)}  If the blocks of a $(q,k)$-PB have at least one element $a$ in common,  
then all blocks are equal, i.e., $B_i=\{a\}$ for all $0\le i<q$.\\
{\rm (iii)}  There is no $(q,k,k+1)$-PB for any prime number $q$ and $k\ge 1$. 
In particular, there is no $(q,1,2)$-PB and there is no $(q,2,3)$-PB.
Also, $\k(3)=1$.
\end{prop}

Construction of $(q,k,t)$-product blocks is not trivial if $k\ge 2$.
Suppose that $k=2$. An interesting question is to find a smallest number $t$ such that 
$(q,2,t)$-product blocks exist for any prime power $q$.
By Proposition \ref{l1}(iii), this number is at least four.
We conjecture that it is at least five for $q\ge 5$.

\begin{con}
A $(q,2,4)$-product blocks do not exist for a prime power $q\ge 5$.
\end{con}

\begin{table} [htb] 
\centering
\begin{tabular}{|c| >{\ttfamily}l| >{\ttfamily}l|}
\hline
$q$ & \multicolumn{1}{|c|}{$(q,2,5)$-product blocks} & *\\
\hline
\hline
5 &
A D B E C & \\ \hline
31 & 
C A E C B E C B E C \ B E C B E C B A D B & \\ & 
A D B A D C A D C A \ D & \\ \hline
37 & 
C B A D C B E C B A \ E C B A D C B E D C & F \\ & 
A E C B A D C B F D \ C B E C B A D & \\ \hline
41 &
B E C B E C A D C F \ E C A D C A D C A D &GH \\ &
B A D B A D B E H B \ A D B E C B E C B E C & \\ \hline
47 &
B E C B A D B A E C \ B E C B E D C A D C  & I \\ &
B E C B A D C B E C \ B A D B A D C A D C &\\ &
B E D I A D C & \\ \hline
53 &
B A D C B E D C B E \ C B A D C B A D C B & HJ
\\ &
E D C B E D C A E C \ B A J C B A D C B E &
\\ &
D C B E H B A D C B \ A J C &
\\ \hline

61 &
C A D B A D B E C B \ E C A D C A D B E D \ \ & J\\ & 
B E C A J C A D B A \ D B E C B E C A D C \ \ & \\ & 
A D B E C B E C A J \ C A D B E D B E C B E &\\ \hline
	
67 &
B E C A E C B E C B \ E C B E C B E C B E \ \ & HIJ\\ & 
C B E D B E C B E H \ B E C B A D B E D B \ \ & \\ & 
A D B A D C A D B A \ D C A D I A J C A D \ \ & \\ & 
C A D C A D C &\\ \hline
71 &
C B E C A D B A D B \ E C A D C A D B E C \ \ & GH\\ & 
B E C A D C A D B E \ C B E C A D B A D B \ \ & \\ & 
E C G E C A D B E H \ B E C A D C A D B E \ \ & \\ & 
C B E C A D B A D B \ E &\\ \hline
73 &
E C B A D C A D C B \ E C B F D C A E C B \ \ & FHJ \\ &
E C B E H B A J C B \ E C B E C B A D C B \ \ & \\ &
E C B E D B A D C A \ E C B F D B A D C A \ \ & \\ & 
D C B E D B A D C A \ D C B &\\ \hline

\end{tabular}
\caption{$(q,2,5)$-product blocks for $q\le 73$.}
\label{tb5a}
\end{table}

\section{$(q,2,t)$-product blocks for $t=5,6$}

In this section we study product blocks for block size $k=2$ using $t=5$ symbols. 
We implemented a program for searching $(q,2,5)$-product blocks using random blocks using elements $\{0,1,2,3,4\}$.
It turns out that in many cases the blocks are 
$A=\{0,1\}, B=\{1,2\}, C=\{2,3\}, D=\{3,4\}, E=\{0,4\}$.
For example, the blocks for $q=5$ are A, D, B, E, C.

The search program did not find $(q,2,5)$-product blocks for any prime $q, 5<q<31$. 
However, it found $(31,2,5)$-product blocks of the following structure.
Only blocks of type $A,B,C,D$, and $E$ are used. 
So, the blocks are represented as a sequence of 31 letters, see Table \ref{tb5a}. 

Tables \ref{tb5a} and \ref{tb5b} show $(q,2,5)$-product blocks for $q<90$ computed using the search program.
The notation in Tables \ref{tb5a} and \ref{tb5b} is the following.
Blocks are labeled as $A=\{0,1\}, B=\{1,2\}, C=\{2,3\}, D=\{3,4\}, E=\{0,4\}$.
Additional blocks are labeled as $F=\{1,4\}, G=\{0,2\}, H=\{2,4\}, I=\{1,3\}, J=\{0,3\}$.
A space is added after every 10 blocks in the sequences.
The column * contains labels $F,G,H,I,J$ used in the corresponding blocks.

\begin{table} [htb] 
\centering
\begin{tabular}{|c| >{\ttfamily}l| >{\ttfamily}l|}
\hline
$q$ & \multicolumn{1}{|c|}{$(q,2,5)$-product blocks} & *\\
\hline
\hline

79 & 
A D C B E D C B F D \ B A D C B E C B A J \ \ & FHJ \\ & 
C B A D C B E C B A \ D C B E D C B E C B \ \ & \\ & 
E H B A D C B A E C \ B E C B A D C B E D \ \ & \\ & 
C B E C B A D C B E \ C B A D C B A D C &\\ \hline
83 & 
A D C A J C A D C B \ E C B E D B A D C A \ \ & HJ \\ &
D C A E C B E C B A \ D B A D B A D C A D \ \ & \\ &
C B E C B E D B E C \ B A D C A D C B E C \ \ & \\ &
B E C B A D B A D C \ A E C B E C B E C B \ \ & \\ &
E H B &\\ \hline
89 &
D B A D C B E C B A \ D C A E C B E C B A \ \ & IJ\\ &
D C B E D I A D C B \ E C B A D I A D C B \ \ & \\ &
A D B A J C B E C B \ A D C B E C B E D C \ \ & \\ &
A D C B E C B A D C \ B E C B A D B A E C \ \ & \\ &
B E D C B E C B E &\\ \hline

\end{tabular}
\caption{$(q,2,5)$-product blocks for $73<q\le 89$.}
\label{tb5b}
\end{table}

When the search program could not find $(q,2,5)$-product blocks for some value of $q$,
it tries to find $(q,2,6)$-product blocks. In some cases only blocks of type $\{i,i+1\}$ 
were found. For example, $(13,2,6)$-product blocks form a sequence 
\[ C, F, C, F, C, F, C, A, D, A, E, B, E,\] where 
$A=\{0,1\}, B=\{1,2\}, C=\{2,3\}, D=\{3,4\}, E=\{4,5\}$, and $F=\{0,5\}$.
We tested all prime numbers up to 100 and found that 
$(q,2,5)$-product blocks exist for
\[ q=5,31,37,41,47,53,61,67,71,73,79,83,89,97\]
and $(q,2,6)$-product blocks exist for 
\[ q= 7,11,13,17,19,23,29,43,59.\] 
 
\begin{table} [htb] 
{\small
\centering
\begin{tabular}{|c|c|l| }
\hline
$q$ & $k$ & \multicolumn{1}{|c|}{$(q,k)$-product blocks} \\
\hline
\hline

5 & 2 & 0 3,  0 2,  2 4,  1 4,  1 3 \\\hline
7 & 2 & 0 3,  2 6,  1 5,  0 4,  3 6,  2 5,  1 4, \\\hline
11 & 3 & 1 6 9,  3 6 9,  0 3 6,  0 3 8,  0 5 8,  2 5 8,  2 5 10,  2 7 10,  4 7 10,  1 4 7, 
\\& &
 1 4 9 \\\hline
17 & 3 &  3 6 11,  3 8 13,  10 11 13,  1 7 10,  1 10 12,  5 8 15,  0 5 12,  9 10 12,  
\\ & &
2 3 12, 7 9 14,  0 2 7,  4 10 14,  1 7 14,  4 11 16,  1 2 8,  1 4 16,  8 13 15\\\hline
19 & 4 &  0 1 8 12,  4 5 12 16,  7 8 15 16,  0 1 11 12,  3 4 8 16,  0 7 8 18,  3 4 10
\\ & &
 15,  7 8 14 15,  0 10 11 18,  2 3 14 15,  6 7 17 18,  2 3 9 10,  2 6 13 14, 
\\ & &
 9 10 17 18,  2 9 13 14,  5 6 16 17,  1 2 8 9,  1 5 12 13,  5 8 9 16\\\hline

23 & 4 &  7 8 12 13,  1 6 17 20,  6 10 13 14,  3 8 19 22,  1 8 12 15,  1 5 10 17,  \\ & &
3 10 14 22,  3 7 8 12,  1 5 19 20,  5 10 14 17,  0 3 21 22,  7 15 16 19,  \\ & &
0 1 6 19,  14 17 18 21,  3 7 11 21,  0 16 19 20,  5 9 13 16,  2 18 21 22,  \\ & &
11 14 15 18,  0 4 8 20,  13 16 17 20,  2 6 10 15,  15 18 19 22
\\\hline

29 & 5 &  7 10 13 16 19,  15 18 21 24 27,  0 3 6 23 26,  2 5 8 11 14,  10 13 16 19 \\ & &
22,  1 18 21 24 27,  0 3 6 9 26,  5 8 11 14 17,  13 16 19 22 25,  1 4 21 \\ & &
24 27,  0 3 6 9 12,  8 11 14 17 20,  16 19 22 25 28,  1 4 7 24 27,  \\ & &
3 6 9 12 15,  11 14 17 20 23,  2 19 22 25 28,  1 4 7 10 27,  6 9 12 15 18,  \\ & &
14 17 20 23 26,  2 5 22 25 28,  1 4 7 10 13,  9 12 15 18 21,  0 17 20 23 26,\\ & &
 2 5 8 25 28,  4 7 10 13 16,  12 15 18 21 24,  0 3 20 23 26,  2 5 8 11 28
\\\hline

31 & 4 &  0 6 7 8,  12 14 22 24,  0 8 18 29,  2 12 14 22,  5 27 29 30,  3 9 11 23,  \\ & &
17 25 27 28,  1 3 11 13,  17 18 25 28,  1 3 5 22,  7 16 18 26,  0 3 10 12,  \\ & &
14 16 17 30,  2 10 14 21,  6 7 14 25,  2 9 23 28,  5 15 16 17,  20 21 23 30,  \\ & &
3 5 6 17,  13 20 22 30,  5 12 26 28,  1 8 19 20,  4 5 23 26,  9 10 20 29,  \\ & &
2 6 23 26,  10 18 19 21,  4 15 24 25,  7 8 9 11,  13 15 22 25,  7 9 11 30,\\ & &
  17 20 22 24
\\\hline

37 & 4 &  3 13 16 31,  18 22 30 33,  14 24 25 30,  6 10 17 35,  8 9 15 30,  1 15\\ & &
20 32,  6 17 18 26,  4 11 30 35,  15 23 28 33,  1 13 18 28,  4 11 13 27,\\ & &
8 13 34 35,  15 21 25 27,  7 8 12 31,  10 12 29 30,  6 10 17 29,  2 3 9 25,\\ & &
9 21 23 34,  2 26 30 33,  10 19 21 36,  1 17 27 33,  5 13 20 36,  5 18 29\\ & &
32,  14 16 22 29,  7 26 31 32,  3 7 24 31,  1 12 22 24,  8 9 18 31,  5 14\\ & &
16 27,  3 5 7 24,  5 7 12 14,  3 12 32 35,  0 11 15 17,  9 13 23 34,\\ & &
 11 13 17 32,  0 9 10 35,  9 20 33 35
\\\hline

\end{tabular}
\caption 
{$(q,k)$-product blocks.}
\label{tbk}
}
\end{table}

\section{$(q,k)$-product blocks}

We develop another search program for computing lower bounds for $\k(q)$ using random blocks of size larger than two.
The $(q,k)$-product blocks for $q<40$ and the best computed values of $k$ are shown in Table \ref{tbk}.
The search program did not find $(13,3)$-product blocks, so the best lower bound for $\k(13)$ is two and the 
corresponding $(13,2)$-product blocks are shown in the previous section. 

An extensive search was done for $q<400$ and the results are shown in Table \ref{tb:pbd}. 
The third column shows the lower bounds of $M(q,q-3)$ computed using Theorem \ref{thm1}.
Based on the results we conjecture the following bounds.

\begin{con}
For all prime numbers $q\ge 19$, $\k(q)\ge 4$.
\end{con}

\begin{con}
For all prime numbers $q\ge 59$, $\k(q)\ge 5$.
\end{con}

\begin{con}
For all prime numbers $q\ge 163$, $\k(q)\ge 6$.
\end{con}

\begin{con}
For all prime numbers $q\ge 293$, $\k(q)\ge 7$.
\end{con}

Notice that $i$th conjecture follows from $(i+1)$st conjecture for $i=2,3$ and 4.

\begin{table} [h!]
\centering
\vspace*{2mm}
\begin{tabular}{|r r r | r r r | r r r|}
\hline
$q$ & $k$ & $M(q,q-3)$ & $q$ & $k$ & $M(q,q-3)$ & 
$q$ & $k$ & $M(q,q-3)$ \\
\hline\hline

5 & 2 & - & 109 & 5 & 70,632 & 251 & 6 & -  \\
7 & 2 & 126 & 113 & 5 & - & 257 & 6 & -  \\
11 & 3 & - & 127 & 5 & 96,012 & 263 & 6 & -  \\
13 & 2 & 468 & 131 & 5 & - & 269 & 6 & -  \\
17 & 3 & - & 137 & 5 & - & 271 & 6 & 512,190  \\
19 & 4 & 1,710 & 139 & 5 & 115,092 & 277 & 6 & 535,164  \\
23 & 4 & - & 149 & 5 & - & 281 & 6 & -  \\
29 & 5 & - & 151 & 5 & 135,900 & 283 & 6 & 558,642  \\
31 & 4 & 4,650 & 157 & 5 & 146,952 & 293 & 7 & -  \\
37 & 4 & 6,660 & 163 & 6 & 184,842 & 307 & 7 & 751,536  \\
41 & 4 & - & 167 & 6 & - & 311 & 7 & -  \\
43 & 4 & 9,030 & 173 & 6 & - & 313 & 7 & 781,248  \\
47 & 4 & - & 179 & 6 & - & 317 & 7 & -  \\
53 & 4 & - & 181 & 6 & 228,060 & 331 & 7 & 873,840  \\
59 & 5 & - & 191 & 6 & - & 337 & 7 & 905,856  \\
61 & 5 & 21,960 & 193 & 6 & 259,392 & 347 & 7 & -  \\
67 & 5 & 26,532 & 197 & 6 & - & 349 & 7 & 971,616  \\
71 & 5 & - & 199 & 6 & 275,814 & 353 & 7 & -  \\
73 & 5 & 31,536 & 211 & 6 & 310,170 & 359 & 7 & -  \\
79 & 5 & 36,972 & 223 & 6 & 346,542 & 367 & 7 & 1,074,576  \\
83 & 5 & - & 227 & 6 & - & 373 & 7 & 1,110,048  \\
89 & 5 & - & 229 & 6 & 365,484 & 379 & 7 & 1,146,096  \\
97 & 5 & 55,872 & 233 & 6 & - & 383 & 7 & -  \\
101 & 5 & - & 239 & 6 & - & 389 & 7 & -  \\
103 & 5 & 63,036 & 241 & 6 & 404,880 & 397 & 7 & 1,257,696  \\
107 & 5 & - & &&&&& \\

\hline
\end{tabular}
\caption{Lower bounds for $\k(q)$ and $M(q,q-3)$ for prime numbers $q<400$.}
\label{tb:pbd}
\end{table}

\end{document}